\documentclass[twocolumn,aps,prd,amsmath,amssymb]{revtex4}

\bibpunct{}{}{,}{s}{}{,}
\begin{document}

\title{Cosmological perturbations through the big bang}

\author{David Wands}
\email{david.wands@port.ac.uk} \affiliation{Institue of Cosmology and Gravitation, University of Portsmouth, Mercantile House, Portsmouth, PO1 2EG, United Kingdom}

\begin{abstract}

Several scenarios have been proposed in which primordial perturbations could originate from quantum vacuum fluctuations in a phase corresponding to a collapse phase (in an Einstein frame) preceding the Big Bang. I briefly review three models which could produce scale-invariant spectra during collapse: (1) curvature perturbations during pressureless collapse, (2) axion field perturbations in a pre big bang scenario, and (3) tachyonic fields during multiple-field ekpyrotic collapse. In the separate universes picture one can derive generalised perturbation equations to describe the evolution of large scale perturbations through a semi-classical bounce, assuming a large-scale limit in which inhomogeneous perturbations can be described by locally homogeneous patches. For adiabatic perturbations there exists a conserved curvature perturbation on large scales, but isocurvature perturbations can change the curvature perturbation through the non-adiabatic pressure perturbation on large scales. Different models for the origin of large scale structure lead to different observational predictions, including gravitational waves and non-Gaussianity.

\bigskip

\center{ {\em to appear in Advanced Science Letters, special issue
on Quantum Gravity, Cosmology and Black Holes} }
\end{abstract}

\maketitle

\section{Introduction}

How did the universe begin? The standard Hot Big Bang model, based
on four-dimension Friedmann-Robertson-Walker (FRW) cosmology, starts
with an initial singularity where our notion of spacetime described
by Einstein's general theory of relativity breaks down. But we do
not expect general relativity, or any classical theory of spacetime,
to hold right up to a Big Bang singularity. Quantum fluctuations
about a simple FRW metric in general relativity, including
first-order inhomogeneities in the geometry in a semi-classical
description, become large as the energy density, and thus the
cosmological expansion rate $H$, becomes comparable to the Planck
scale, $10^{19}$~GeV. In alternative models, such as models with
large extra dimensions, the classical four-dimensional effective
theory may break down at much lower energies.

Cosmology can be studied without worrying about what came before
the Big Bang as long as we have some prescription for the initial
conditions, at whatever time we choose to apply the rules of general
relativity, or some model of four-dimensional semi-classical
gravity. In the homogeneous and isotropic FRW cosmology it may be
sufficient to specify an initial thermal temperature and evolve this
forward to the present day. But the standard Hot Big Bang model does
not give a unique prescription for the initial distribution of
inhomogeneities - spatial variations in the matter and geometry
across the initial spatial hypersurface. Indeed there is no reason
that they should necessarily be small perturbations, but
observations (notably of the cosmic microwave background) suggest
they are.

There are two logical possibilities for the origin of primordial
perturbations. Either they are produced after the Big Bang, or they
originate before the Big Bang.

There is a simple model to generate an
initial spectrum primordial perturbations due to vacuum fluctuations
during inflation driven by a slowly-rolling,
self-interacting scalar field. The accelerated expansion leads to
the vacuum fluctuations on small scales (much smaller than the
Hubble length, $H^{-1}$) being swept up to large (super-Hubble)
scales where they become ``frozen-in'' by the cosmological
expansion. The amplitude of the fluctuations at the Hubble scale is
proportional to $H$ and the slowly varying expansion rate leads to
an almost scale invariant spectrum. The weak interactions required
for a slowly-rolling field naturally lead to an almost Gaussian
distribution for the primordial perturbations on large scales. After
almost 30 years of theoretical development this inflationary picture
of the early universe has become the standard model for the origin
of structure \cite{Liddle:2000cg}. There is no single agreed model for which fundamental
field is responsible for driving inflation and/or generating
structure, but there are numerous possibilities based on
extensions beyond the standard model of particle physics \cite{Lyth:1998xn}.

But it is also possible that the large scale structure of our
Universe is inherited from vacuum fluctuations during an earlier
non-inflationary phase, {\em before} the Big Bang\cite{Gasperini:1992em}. It is this
possibility that I will discuss in this paper. There are many
similarities with the inflationary model for the origin of structure
in that one can calculate a spectrum of perturbations on large,
super-Hubble scales in the Hot Big Bang model assuming only vacuum
fluctuations on small, sub-Hubble scales in a preceding phase, only
one now assumes that the preceding phase was one of accelerated
contraction (in the Einstein frame where general relativity
applies). This requires an intermediate bounce from contraction to
expansion and one of the unresolved problems is whether such a
bounce really occurs in this manner and if so whether the
perturbations spectrum calculated in the collapse phase can
be related to perturbations in the standard Hot Big Bang. I will
argue that under fairly general conditions the spectrum on large
scales of interest can be expected to be preserved through a bounce,
while acknowledging that there is as yet no entirely satisfactory
physical model for the bounce.

The existence of a cosmological phase before the Big Bang leads to a radically different view regarding the initial conditions for the universe \cite{Buonanno:1998bi,Steinhardt:2001st}, and such models have been criticised \cite{Turner:1997ih,Linde:2002ws} for requiring a very large universe (relative to the Planck scale) before the big bang. Indeed, as I will discuss, in some cases the generation of a scale-invariant spectrum of primordial perturbations requires the unstable growth of perturbations during collapse. However given the uncertainty in what constitutes a ``natural'' initial state for the cosmos, I will consider the possible observational consequences of a collapse era.

\section{Homogeneous collapse}

In this paper I will consider a four-dimensional background
cosmological model that is spatially homogeneous and isotropic and
therefore described by the Friedmann-Robertson-Walker metric
\begin{equation}
ds^2 = -dt^2 + a^2(t) \gamma_{ij} dx^i dx^j \,.
\end{equation}
where $\gamma_{ij}$ is the metric on a maximally symmetric 3-space
with uniform curvature $K$. The Hubble expansion rate (or collapse
rate) is $H\equiv \dot{a}/a$.

Local energy conservation gives the continuity equation for matter
 \begin{equation}
  \dot\rho = -3H(\rho+P) \,,
 \end{equation}
where $\rho$ is the energy density and $P$ the isotropic pressure.
For a linear barotropic equation of state $P=w\rho$ this can be
integrated to give $\rho \propto a^{-3(1+w)}$.

In a collapsing universe, $\dot{a}<0$, in the presence of matter
with $\rho+3P>0$ (or $w>-1/3$) the energy density grows faster
than the spatial curvature $K/a^2$. Note that in an expanding
universe one requires $\rho+3P<0$ (or $w<-1/3$) for the energy
density to grow relative to the spatial density, and this is the
usual condition for inflation. For simplicity I will assume in the
following that spatial curvature is negligible, so that
$\gamma_{ij}=\delta_{ij}$.
More problematic in a collapsing universe is anisotropic shear\cite{Kunze:1999xp}.
In the simplest case of a Bianchi I universe the shear is proportional
to $a^{-6}$ (where in this case we can still think of $a^3$ as the
volume factor). Thus the anisotropic shear grows relative to matter
in a collapsing universe for any matter with $P<\rho$.

Although a fluid description yields simple linear barotropic
equation of state (for matter $w=0$, or radiation $w=1/3$) I will be
interested in microphysical description of the matter where one can
use a quantum vacuum state to set the initial conditions for
inhomogeneous perturbations at early times. Thus I will consider
canonical scalar fields $\varphi_I$ with energy density and pressure
 \begin{eqnarray}
 \rho = V(\varphi_I) + \sum_I \frac12 \dot\varphi_I^2 \,,\\
 P = - V(\varphi_I) + \sum_I \frac12 \dot\varphi_I^2 \,.
 \end{eqnarray}
where $V(\varphi_I)$ is the potential energy. In particular a scalar
field with an exponential potential, $V_I(\varphi_I)\propto
\exp(-\lambda_I\kappa\varphi_I)$ where $\kappa^2=8\pi G$, provides a simple model with $P=w\rho$
where $1+w=\lambda^2/3$. This is the basis of both power-law
inflation \cite{Lucchin:1984yf} for $\lambda^2<2$ and ekpyrotic
collapse \cite{Khoury:2001wf} with $\lambda^2\gg 2$.

Canonical scalar fields also have a kinetic-dominated cosmology if
the potential energy can be neglected such that $P=\rho=\sum_I
\dot\varphi_I^2/2$, corresponding to a stiff equation of state with
$w=1$. Indeed in a collapsing universe where the energy density
grows as the universe collapses, the kinetic energy eventually
dominates over any finite potential energy.

We can identify three scalar-field collapse scenarios based on the
form of the potential \cite{Heard:2002dr,Erickson:2003zm}:
\begin{itemize}
\item
Non-stiff collapse with $P<\rho$: stable with respect to spatial
curvature for $P>-\rho/3$ but unstable with respect to anisotropic
shear.
\item
Pre-Big Bang \cite{Gasperini:1992em} collapse $P=\rho$: stable with
respect to spatial curvature and marginally stable with respect to
anisotropic shear\cite{Kunze:1999xp}.
\item
Ekpyrotic collapse \cite{Khoury:2001wf} with $P\gg \rho$: stable
with respect to spatial curvature and anisotropic shear.
\end{itemize}

The last two are models which have been inspired by ideas from
string theory and are intrinsically higher-dimensional models.
Nonetheless most of the quantitative results have been developed for
effective theories describing scalar fields in four-dimensional
spacetime. Both take as their starting point the notion that string
theory should be a self-consistent theory without the singularities
found in general relativity.

The pre Big Bang model
\cite{Gasperini:1992em,Lidsey:1999mc,Gasperini:2002bn} assumes that
our observable Universe began in a low energy, weak coupling state
well described by a low energy effective action of string theory.
Although sometimes written in terms of an expanding cosmology in the
string frame where the dilaton field is non-minimally coupled to the
spacetime curvature, this solution can be conformally transformed to
a collapsing cosmology described by general relativity in the
so-called Einstein frame \cite{Gasperini:1993hu}. As the dynamics is
dominated by the kinetic energy it is independent of the form of the
potential or even the number of fields. As the energy density
becomes large, and the dilaton becomes large, the low-energy and
weak-coupling approximations inevitably break down. This offers the
possibility that the general relativistic singularity is resolved,
but this goes beyond the low-energy effective description\cite{Brustein:1997cv,Cartier:1999vk}.

The ekpyrotic model
\cite{Khoury:2001wf,Kallosh:2001ai,Khoury:2001iy} was originally
motivated by cosmological solutions describing the motion of branes
in a higher-dimensional spacetime. But again this is usually
described by an effective theory of scalar fields in a
four-dimensional spacetime. In contrast to the pre Big Bang, it
incorporates a negative effective potential which is unbounded from
below, leading to approximately power-law collapse model with
$w\gg1$. This ultra-stiff fast-roll collapse driven by a steep,
negative potential is in many ways dual to quasi-de Sitter,
slow-roll inflation driven by a flat, positive potential. Unlike the
pre Big Bang the model approaches the weak coupling during the
collapse phase. In the original model the authors appealed to the
higher-dimensional picture to resolve the apparent singularity in
the four-dimensional effective theory \cite{Khoury:2001bz}.

\section{Linear perturbations during collapse}

\subsection{Free field perturbations in an FRW cosmology}

Let us first consider the dynamics of free field perturbations in a
FRW background with scale factor $a$ and Hubble
rate, $H\equiv \dot{a}/a$, where a dot denotes derivatives with
respect to cosmic time $t$.

Consider an inhomogeneous perturbation,
$\varphi_I\to\varphi_I(t)+\delta\varphi_I(t,{\bf x})$, of the
Klein-Gordon equation for a scalar field in an unperturbed FRW
universe:
 \begin{equation}
\label{simpleKG} \ddot{\delta\varphi}_I +3H\dot{\delta\varphi}_I +
\left( m_I^2 - \nabla^2 \right) \delta\varphi_I = 0 \,,
 \end{equation}
where the effective mass-squared of the field is
$m_I^2=\partial^2V/\partial\varphi_I^2$, and $\nabla^2$ is the
spatial Laplacian. Decomposing an arbitrary field perturbation into
eigenmodes of the spatial Laplacian (Fourier modes in flat space)
$\nabla^2\delta\varphi_I=-(k^2/a^2)\delta\varphi_I$, where $k$ is
the comoving wavenumber, we find that
small-scale fluctuations undergo underdamped oscillations on
sub-Hubble scales (with comoving wavenumber $k>aH$), but on large
super-Hubble scales, $k<aH$, the modes are overdamped (or
``frozen-in'').

This is most clearly seen in terms of the rescaled field and
conformal time
\begin{equation}
 v_I = a \delta\varphi_I \,, \quad
 a d\eta = dt \,.
 \end{equation}
The Klein-Gordon equation (\ref{simpleKG}) becomes
\begin{equation}
 \label{KGv}
 v_I'' + \left( k^2 + a^2 m_I^2 - \frac{a''}{a} \right) v_I = 0 \,.
 \end{equation}
We have a simple harmonic oscillator with time-dependent effective
mass
\begin{equation}
 \label{mu}
 \mu^2 = m_I^2 a^2 - \frac{a''}{a} = \left( m^2 - \frac{1-3w}{2} H^2 \right) a^2 \,.
\end{equation}

For $k^2/a^2\gg |1-3w|H^2$ and $k^2/a^2\gg |m^2|$ we can neglect the
effective mass and we have essentially free oscillations.
Normalising the initial amplitude of these small-scale fluctuations
to the zero-point fluctuations of a free field in flat spacetime we
have \cite{LLKCreview}
\begin{equation}
 \label{vacuumdeltaphi}
\delta\varphi_I \simeq \frac{e^{-ikt/a}}{a\sqrt{2k}} \,.
\end{equation}

During an accelerated expansion or collapse $|\dot{a}|=|aH|$
increases and modes that start on sub-Hubble scales ($k^2>a^2H^2$)
are stretched up to super-Hubble scales ($k^2<a^2H^2$). For $k^2\ll
|1-3w|a^2H^2$ we can neglect the spatial gradients in
Eq.~(\ref{simpleKG}). Perturbations in light fields (with
mass-squared $m^2\ll |1-3w|H^2$) become over-damped (or
``frozen-in'') and Eq.~(\ref{vacuumdeltaphi}) evaluated when
$k\simeq aH$ gives the power spectrum for scalar field fluctuations
at ``Hubble-exit''
\begin{equation}
\label{H2pi}
 \left. {\cal P}_{\delta\varphi_I} \right|_{k=aH} \equiv \frac{4\pi
k^3}{(2\pi)^3} \left| \delta\varphi_I^2 \right|_{k=aH} \simeq \left(
\frac{H}{2\pi} \right)^2 \,.
\end{equation}
Heavy fields with $m^2\gg |1-3w|H^2/4$ remain under-damped and have
essentially no perturbations on super-Hubble scales. But light
fields become over-damped and can be treated as essentially
classical perturbations with a Gaussian distribution on super-Hubble
scales. Then on large scales we have
\begin{equation}
 \delta\varphi_I \simeq C + D \int \frac{dt}{a^3(t)} \,.
 \end{equation}
where $C$ and $D$ are constants of integration.
In an expanding universe with $w<1$ the integral converges and the
field fluctuations become frozen-in on large scales. However in a
collapsing universe with $w<1$, there is a growing mode at late
times. This is due to the instability with respect to the kinetic
energy of the field which (like anisotropic shear) grows as
$\dot{\delta\varphi}^2\propto a^{-6}$ in a collapsing universe.

Vacuum fluctuations in massless fields during quasi-de Sitter
expansion ($|\dot{H}|\ll H^2$) produces approximately constant
amplitude of scalar field fluctuations at Hubble-exit which then
remain approximately constant on super-Hubble scales, thus producing
an approximately scale-invariant spectrum. During an accelerated
collapse $H^2$ grows rapidly ($\dot{H}=-3(1+w)H^2/2$) and thus the
typical amplitude of fluctuations at Hubble-exit grows rapidly with
time. During pre-big bang or ekpyrotic collapse with $w\geq1$ these
perturbations are frozen-in and hence minimally coupled, massless
fields acquire a steep blue spectrum. On the other hand if $w<1$ the
instability causes perturbations to grow on super-Hubble scales and
in the particular case of a pressureless collapse ($w=0$) the
super-Hubble growth exactly matches the growth of perturbations at
Hubble-exit leading to a scale-invariant spectrum of perturbations
on super-Hubble scales
\cite{Starobinsky:1979ty,Wands:1998yp,Finelli:2001sr}.

It is interesting to note that it is the presence of an instability
that enables the collapse phase with growing $H^2$ to produce a
scale-invariant spectrum \cite{Creminelli:2004jg}. It is also possible to
produce a scale invariant spectrum if the field has a tachyonic
mass, $m^2<0$, which again leads to super-Hubble modes that grow at
precisely the same rate as the fluctuations at Hubble-exit
\cite{Finelli:2002we}. Another means to produce a scale-invariant
spectrum is due to a non-minimal coupling, as in the case of
pseudo-scalar axion fields in the pre big bang
scenario~\cite{Copeland:1997ug}.


Thus far we have neglected the interactions of the field
perturbations, including the gravitational coupling, which is valid
only for isocurvature field perturbations, whose energy-momentum is
negligible. In the next section we will include the effect of linear
metric perturbations.

\subsection{Scalar field and metric perturbations, with interactions}

To track the evolution of more general perturbations we need to
include interactions between fields and, even in the absence of
explicit interactions, we need to include gravitational coupling via
metric perturbations.


For an inhomogeneous matter distribution the Einstein equations
imply that we must also consider inhomogeneous metric perturbations
about the spatially flat FRW metric. The perturbed FRW spacetime is
described by the line element~\cite{Mukhanov90}
\begin{eqnarray}
 \label{metric}
\hspace*{-0.2em} ds^2 &=& - (1+2A) dt^2 + 2a\partial_iB dx^i dt
\nonumber\\
\hspace*{-0.2em}&& +a^2\left[ (1-2\psi)\delta_{ij} + 2\partial_{ij}E
 + h_{ij} \right] dx^i dx^j\,,
\end{eqnarray}
where $\partial_i$ denotes the spatial partial derivative
$\partial/\partial x^i$. We will use lower case latin indices to run
over the 3 spatial coordinates.

The metric perturbations have been split into scalar and tensor
parts according to their transformation properties on the spatial
hypersurfaces. The field equations for the scalar and tensor parts
then decouple to linear order. Vector metric perturbations are
related to the divergence-free part of the momentum, which vanishes
identically at linear order for minimally coupled scalar fields.
However vector perturbations have been studied, for example, in the
pre big bang model as possible source of primordial magnetic fields
due to the non-minimal coupling of the dilaton field
\cite{Lemoine:1995vj}.


The tensor perturbations, $h_{ij}$, are transverse ($\partial^i
h_{ij}=0$) and trace-free ($\delta^{ij}h_{ij}=0$). They are
automatically independent of coordinate gauge transformations.
These describe gravitational waves as they are the free part of the
gravitational field and evolve independently of linear matter
perturbations.

We can decompose arbitrary tensor perturbations into eigenmodes of
the spatial Laplacian, $\nabla^2e_{ij}=-(k^2/a^2)e^{(+,\times)}_{ij}$, with two possible polarisation states, $+$ and $\times$, comoving wavenumber $k$, and scalar amplitude $h(t)$:
\begin{equation}
\label{eq:defh} h_{ij} = h(t) e_{ij}^{(+,\times)}(x)\,.
\end{equation}
The Einstein equations yield a wave equation for the amplitude of
the tensor metric perturbations
\begin{equation}
 \label{teneq}
\ddot{h} + 3H\dot{h} + \frac{k^2}{a^2} h = 0 \,,
\end{equation}
This is the same as the wave equation (\ref{simpleKG}) for a
massless scalar field in an unperturbed FRW metric. Thus initial
vacuum fluctuations on sub-Hubble scales give rise to a power
spectrum for tensor metric fluctuations at Hubble-exit \cite{RMP}
proportional to that given in Eq.~(\ref{H2pi})
\begin{equation}
\left. {\cal P}_h \right|_{k=ah} = 64\pi G \left( \frac{H}{2\pi}
\right)^2 \,.
\end{equation}
Note that as the Hubble rate approaches the Planck scale, $H^2\to G^{-1}$, the power in metric perturbations becomes of order unity, signalling the expected breakdown of the semi-classical description.


The four scalar metric perturbations $A$, $\partial_iB$,
$\psi\delta_{ij}$ and $\partial_{ij}E$ are constructed from
3-scalars, their derivatives, and the background spatial metric. In
particular the intrinsic Ricci scalar curvature of constant time
hypersurfaces is given by
\begin{equation}
^{(3)}R = \frac{4}{a^2} \nabla^2 \psi \,.
\end{equation}
%


First-order perturbations of a canonical scalar field in a
first-order perturbed FRW universe obey the wave equation~\cite{RMP}
\begin{eqnarray}
\label{eq:pertKG} \label{eq:scalareom} \hspace*{-2.0em}&&
\ddot{\delta\varphi}_I + 3H\dot{\delta\varphi}_I
 + \frac{k^2}{a^2} \delta\varphi_I + \sum_J V_{IJ}
\delta\varphi_J
  \nonumber\\
  \hspace*{-2.0em}&&~~{}
= -2V_{I}A + \dot\varphi_I \left[ \dot{A} + 3\dot{\psi} +
\frac{k^2}{a^2} (a^2\dot{E}-aB) \right]. \label{eq:perturbation}
\end{eqnarray}
where the mass-matrix $V_{IJ}\equiv\partial^2
V/\partial\varphi_I\partial\varphi_J$.
The Einstein equations relate the scalar metric perturbations to
matter perturbations via the energy and momentum constraints
\cite{Mukhanov90}
\begin{eqnarray}
\label{eq:densitycon}
 -4\pi G \delta\rho
 &=&
 3H\left(\dot\psi+HA\right) \nonumber \\
&& + \frac{k^2}{a^2}\left[\psi+H(a^2\dot{E}-aB)\right]
  \,, \\
 -4\pi G \delta q   &=& \dot\psi + HA
\,, \label{eq:mtmcon}
\end{eqnarray}
where the energy and pressure perturbations and momentum for $n$
scalar fields are given by~\cite{RMP}
\begin{eqnarray}
\delta\rho &=& \sum_I\left[
 \dot\varphi_I \left( \dot{\delta\varphi}_I -\dot\varphi_I A \right)
 + V_{I}\delta\varphi_I \right] \,,
\label{eq:densityphi} \\
\delta P &=& \sum_I\left[
 \dot\varphi_I \left( \dot{\delta\varphi}_I -\dot\varphi_I A \right)
 - V_{I}\delta\varphi_I \right] \,,
\label{eq:pressurephi} \\
\delta q_{,i} &=& - \sum_I \dot{\varphi}_I \delta\varphi_{I,i} \,,
\label{eq:mtmphi}
\end{eqnarray}
where $V_I \equiv \partial V/\partial \varphi_I$.


We can construct a variety of gauge-invariant combinations of the
scalar metric perturbations. The longitudinal gauge corresponds to a
specific gauge-transformation to a (zero-shear) frame such that
$E=B=0$, leaving the gauge-invariant variables
\begin{eqnarray}
 \label{defPhi}
\Phi &\equiv& A - \frac{d}{dt} \left[ a^2(\dot{E}-B/a)\right] \,,\\
 \label{defPsi}
\Psi &\equiv& \psi + a^2 H (\dot{E}-B/a) \,.
\end{eqnarray}
%

%

Another variable commonly used to describe scalar perturbations
during inflation is the field perturbation in the spatially flat
gauge (where $\psi=0$). This has the gauge-invariant definition
\cite{Mukhanov85,Sasaki86}:
\begin{equation}
 \label{eq:defdphipsi}
 \delta\varphi_{I\psi} \equiv \delta\varphi_I + \frac{\dot\varphi}{H} \psi \,.
\end{equation}
%
%
It is possible to use the Einstein equations to eliminate the metric
perturbations from the perturbed Klein-Gordon equation
(\ref{eq:pertKG}), and write a wave equation solely in terms of the
field perturbations in the spatially flat gauge
\cite{Sasaki95}
\begin{eqnarray}
   \ddot{\delta\varphi}_{I\psi}
 + 3H\dot{\delta\varphi}_{I\psi}
 + \frac{k^2}{a^2}\delta\varphi_{I\psi}
  \hspace{1.1in}
 \nonumber\\
  +
  \sum_J
   \left[ V_{IJ} -\frac{8\pi G}{a^3}
 \frac{d}{dt} \left( \frac{a^3}{H} \dot{\varphi}_I\dot{\varphi}_J
 \right) \right]
   \delta\varphi_{J\psi}=0
 \,.
 \end{eqnarray}
This generalises the single free-field Klein-Gordon equation
(\ref{simpleKG}) to multiple, interacting fields.

For any light fields (whose masses are small compared to the Hubble
scale) the amplitude of perturbations at Hubble-exit is
approximately given by Eq.~(\ref{H2pi}), however the evolution on
large, super-Hubble scales now depends on the fields interactions.


It is often useful to identify the ``adiabatic'' field perturbation
which is a perturbation forwards or backwards along the background
trajectory in field space \cite{Gordon:2000hv} (see van Tent
\cite{vanTent} for the generalisation to non-canonical fields)
\begin{equation}
 \label{adiabaticpert}
 \delta\sigma = \sum_I \frac{\dot\varphi_I \delta\varphi_I}{\dot\sigma^2} \,,
\end{equation}
where
\begin{equation}
 \dot\sigma^2 \equiv \sum_I \dot\varphi_I^2 \,.
\end{equation}
The total energy, pressure and momentum perturbations for multiple
fields in Eqs.~(\ref{eq:densityphi}--\ref{eq:mtmphi}) can be written
as \cite{RMP}
\begin{eqnarray}
 \delta\rho &=& \dot\sigma \left( \dot{\delta\sigma} -\dot\sigma A \right)
 + V_\sigma\delta\sigma + V_s\delta s \,,
\label{eq:addensityphi} \\
\delta P &=& \dot\sigma \left( \dot{\delta\sigma} -\dot\sigma A
\right)
 - V_\sigma\delta\sigma \,,
\label{eq:adpressurephi} \\
\delta q_{,i} &=& - \dot\sigma \delta\sigma_{,i} \,,
\label{eq:admtmphi}
\end{eqnarray}
where $V_\sigma\equiv (\partial V/\partial
\varphi_I)\dot\varphi_I/\dot\sigma$. The only effect of isocurvature
field perturbations, orthogonal to the background trajectory, is
through a non-adiabatic pressure perturbation
\begin{eqnarray}
 P_{{\rm nad},s} &=& -2 \delta_s V \nonumber\\
 &=& 2 \left(  V_\sigma\delta\sigma - \sum_I V_I \delta\varphi_I \right) \,.
\end{eqnarray}

If the potential gradients vanish orthogonal to the background
trajectory, $\delta_s V=0$ (for isocurvature fields at a local
extremum of their potential), then the adiabatic and isocurvature
field perturbations decouple. The isocurvature perturbations obey
the Klein-Gordon equation (\ref{simpleKG}) while the adiabatic field
perturbations (on spatially flat hypersurfaces) obey the
Klein-Gordon equation for a single field in a perturbed FRW
cosmology
\begin{equation}
    \ddot{\delta\sigma}_{\psi}
 + 3H\dot{\delta\sigma}_{\psi}
 + \left[ \frac{k^2}{a^2} + V_{\sigma\sigma} -\frac{8\pi G}{a^3}
 \frac{d}{dt} \left( \frac{a^3}{H} \dot\sigma^2 \right) \right]
   \delta\sigma_{\psi} = 0
 \,.
\end{equation}
where the final term on the left-hand-side describes the
gravitational back-reaction due to metric perturbations. This is
compactly written in terms of conformal time and $v\equiv
a\delta\sigma_\psi$ and $z\equiv a\dot\sigma/H$ to give
\begin{equation}
 \label{KGvz}
 v'' + \left( k^2 - \frac{z''}{z} \right) v = 0 \,.
\end{equation}
Analogous to Eq.~(\ref{KGv}) this leads to oscillating solutions on
small scales, while on large scales where we neglect the spatial
gradients we obtain
\begin{equation}
 \delta\sigma \simeq \frac{C\dot\sigma}{H} + \frac{D\dot\sigma}{H} \int \frac{H^2 dt}{a^3 \dot\sigma^2} \,.
\end{equation}
In particular we see that the comoving curvature perturbation
\begin{equation}
 {\cal R} \equiv \psi + \frac{H}{\dot\sigma}\delta\sigma = \frac{v}{z} \,,
\end{equation}
has a constant mode on large scales.
During slow-roll inflation the second mode decays rapidly and is
usually neglected,
 but it may become a growing mode during an accelerated collapse.


\section{Three ways to scale-invariant spectra}

\subsection{Collapse with $P\ll\rho$}

If we consider a power-law collapse, $a\propto (-t)^p$ where
$p=2/3(1+w)$, driven by a scalar field with exponential potential
then we have $\dot{H}=-4\pi G\dot\sigma^2= -(1/p)H^2$ and hence
$z''/z=a''/a$ and Eq.~(\ref{KGvz}) for the adiabatic field reduces
to Eq.~(\ref{KGv}) for the isocurvature fields with
\begin{equation}
 \frac{z''}{z} = \frac{a''}{a} = \frac{\nu^2-(1/4)}{\eta^2} \,.
\end{equation}
where
\begin{equation}
 \label{nu}
 \nu = \frac{3}{2} + \frac{1}{p-1} \,.
\end{equation}

The general solution is given in terms of Hankel functions of order
$|\nu|$
\begin{equation}
 \label{Hankel}
 v = \sqrt{|k\eta|} \left[ V_+ H_{|\nu|}^{(1)} (|k\eta|) + V_- H_{|\nu|}^{(2)} (|k\eta|) \right] \,.
\end{equation}
Normalising to the quantum vacuum on sub-Hubble scales at
$\eta\to-\infty$ gives a spectrum of field perturbations on
super-Hubble scales \cite{Wands:1998yp} as $\eta\to0$
\begin{equation}
 \label{Pnu}
 {\cal P}_{\delta\sigma} = \left( \frac{2^{|\nu|}\Gamma(|\nu|)}{(\nu-1/2)2^{3/2}\Gamma(3/2)} \right)^2
 \left( \frac{H}{2\pi} \right)^2 \left| k\eta \right|^{3-2|\nu|} \,.
\end{equation}
Thus a power-law collapse gives rise to a power-law spectrum for
field fluctuations on super-Hubble scales with spectral tilt
\begin{equation}
 \label{ns}
 \Delta n_{\delta\sigma} \equiv \frac{d\ln {\cal P}_{\delta\sigma}}{d\ln k} = 3-2|\nu| \,.
\end{equation}

I have written these expressions in a way that makes clear that the
spectral tilt is invariant under a change of sign of $\nu\to-\nu$,
or equivalently \cite{Wands:1998yp}
\begin{equation}
 \label{dualp}
 p \to \frac{1-2p}{2-3p} \,.
\end{equation}
In particular we see that a scale invariant spectrum of fluctuations
in the adiabatic field ($\Delta n_{\delta\sigma}= 0$) may be
produced either from slow-roll inflation ($w=-1$ and $\nu=3/2$) or a
pressureless collapse \cite{Wands:1998yp,Finelli:2001sr} ($w=0$ and $\nu=-3/2$). This is because the
general solution contains two modes and the transformation
(\ref{dualp}) swaps the growing and decaying modes at late times. In
slow-roll inflation it is the constant mode outside the Hubble-scale
which acquires a scale-invariant spectrum whereas in $w=0$ collapse
it is the time-dependent mode which grows rapidly ${\cal
P}_{\delta\sigma}\propto H^2$ outside the Hubble-scale. This is
evidence of an instability of the background solution describing
pressureless collapse with $\rho\propto a^{-3}$ which is unstable to
the growth of scalar field kinetic energy with
$\dot{\delta\sigma}^2\propto a^{-6}$. This raises questions about
how fine-tuned the initial conditions would need to be to have a
long-lasting, pressureless collapse phase. But if there is such a
phase, even if it is short-lived, then it can generate a
scale-invariant spectrum of perturbations over some range of scales.

Note that the spectrum of adiabatic field fluctuations, massless
isocurvature field fluctuations, and gravitational waves \cite{Starobinsky:1979ty} all share
the same scale-dependence in a power-law collapse. There is a simple
relation between the power of tensor to scalar metric perturbations
during power-law collapse
\begin{equation}
 r \equiv \frac{{\cal P}_h}{{\cal P}_{\cal R}} = \frac{64\pi G \dot\sigma^2}{H^2} = \frac{16}{p} \,.
\end{equation}
This is small for slow-roll inflation, but is dangerously large
($r=8$) during pressureless collapse.  Current observational bounds
\cite{Dunkley:2008ie} require $r<0.3$ at the time of last-scattering
of the CMB. However whereas the tensor and scalar amplitudes are
constant on large scales during conventional slow-roll inflation,
both quantities are rapidly growing during pressureless collapse.
Thus the final value for the tensor-to-scalar ratio will be
model-dependent. In a simple bounce model \cite{Allen:2004vz} it was
found that the tensor-scalar ratio was small in only a small corner
of parameter space.


\subsection{Pre big bang with $P=\rho$}

The pre big bang scenario is based upon bosonic fields in the
low-energy string effective action including the dilaton and other
moduli fields\cite{Gasperini:1992em,Lidsey:1999mc,Gasperini:2002bn}.
Any finite potential becomes negligible as the energy
density grows in a collapsing universe and hence the universe
becomes dominated by the kinetic energy of the fields, leading to
power-law collapse with $w=1$ and $p=1/3$. In this case adiabatic
and canonical isocurvature field perturbations have a general
solution given by Eq.~(\ref{Hankel}) with Hankel functions of order
$\nu=0$. This gives a strong blue spectral tilt $\Delta n=+3$ in
Eq.~(\ref{ns}) for both the scalar and tensor metric perturbations
during the pre big bang phase\cite{Brustein:1994kn}, leaving essentially no perturbations
on large scales.

Originally it was hoped that if the pre big bang could provide a homogeneous universe on large scales then causal mechanisms such as cosmic strings or other
topological defects could source primordial perturbations. However
subsequent observations \cite{Dunkley:2008ie} have shown that an approximately
Gaussian distribution of adiabatic density perturbations is required
on super-Hubble scales by the time of last scattering.

It turns out that some isocurvature fields will have very different
perturbation spectra if they are non-minimally coupled to fields
such as the dilaton which are rapidly evolving during the pre big
bang. In particular the pseudo-scalar axion in the four-dimensional
effective action is coupled to the dilaton in an SL(2,R) invariant
Lagrangian \cite{Lidsey:1999mc}
\begin{equation}
 {\cal L} = - \frac12 (\nabla\sigma)^2 - \frac12 e^{2\sigma} (\nabla\chi)^2 \,.
\end{equation}
The Klein-Gordon equation for isocurvature fluctuations in the axion
field is
\begin{equation}
 \ddot\delta\chi + \left( 3H + 2\dot\sigma \right) \dot\delta\chi + \frac{k^2}{a^2} \delta\chi = 0 \,.
\end{equation}
Analogous to Eq.~(\ref{KGv}) this can be written as
\begin{equation}
 u'' + \left( k^2 - \frac{\bar{a}''}{\bar{a}} \right) u = 0 \,,
\end{equation}
where $u=\bar{a}\delta\chi$ and $\bar{a}\equiv e^{\sigma} a$ is the
``scale factor'' in the conformal frame in which the axion (rather
than the dilaton) is minimally coupled. As a result the spectral
index for axion field perturbations turns out to be given by
\cite{Copeland:1997ug}
\begin{equation}
 \Delta n_{\delta\chi} = 3 - 2|\cos\xi| \,,
\end{equation}
where $\xi$ is an angle describing the rate at which the dilaton
rolls relative to other moduli fields. The invariance of the spectra
under $\cos\xi\to-\cos\xi$ corresponds to the previously noted
invariance under $\nu\to-\nu$ which here coincides with invariance
under duality transformations of the string effective action
\cite{Lidsey:1999mc}. Perturbations of the coupled dilaton-axion system can
be shown to be invariant under SL(2,R) transformations of the
background solutions \cite{Copeland:1997ug}.

More generally there are many axion-type fields in the
four-dimensional effective theories with different couplings to the
dilaton and/or other moduli fields. For specific parameters these
may acquire scale-invariant, or almost scale-invariant spectra \cite{Lidsey:1999mc}.

These isocurvature perturbations during the pre big bang phase still need to be converted into
adiabatic density perturbations in the primordial era. This
will happen if the axion field leads to a non-adiabatic pressure
perturbation, $\delta P_{\rm nad}$, and hence a perturbation in the
local equation of state which changes the large-scale curvature
perturbation
\begin{equation}
 \dot{\cal R} \simeq H \frac{\delta P_{\rm nad}}{\rho+P} \,.
\end{equation}
In recent years a number of such mechanisms have been investigated
in the context of inflationary cosmology with multiple fields.

In the curvaton scenario \cite{Enqvist:2001zp,Lyth:2001nq}, the axion
survives from the pre big bang into the hot big bang phase, as a
massive, weakly coupled field. Although it's initial energy density
is negligible, once it becomes non-relativistic its energy density
grows relative to the radiation and can eventually come to
contribute a significant fraction of the total energy density. The
curvaton must decay before primordial nucleosynthesis, but when it
does so, any perturbation in its energy density is transferred to
the radiation density.  Curvaton models have distinctive
observational signatures including the possibility of residual
isocurvature modes \cite{Lyth:2002my} or non-Gaussianity in the
primordial density perturbation \cite{Lyth:2002my,Sasaki:2006kq}.

Note that the pre big bang is only marginally stable with respect to
anisotropic shear\cite{Kunze:1999xp} in a collapsing universe, and anisotropies grow
during any collapse with $P<\rho$.

\subsection{Ekpyrotic collapse with $P\gg\rho$}

The ekpyrotic scenario involves a collapse phase driven by a steep
and negative exponential potential in the four-dimensional effective
action. This leads to an ultra-stiff equation of state $w\gg1$ and a
rapidly increasing Hubble rate while the scale factor only slowly
decreases. This is in many ways the collapse equivalent of slow-roll
inflation where the scale factor rapidly increases while the Hubble
rate slowly decreases. The ekpyrotic collapse is the stable
attractor during collapse with respect to spatial curvature and
shear, just as slow-roll inflation is the stable attractor during
expansion.

However for $w\gg1$ and thus power-law collapse with $p\ll 1$ the
spectral index given in Eqs.~(\ref{nu}) and~(\ref{ns}) for scalar
and tensor metric perturbations produced during collapse is steep
and blue\cite{Lyth:2001pf} $\Delta n=+2$. This is in contrast to the
original ekpyrotic papers which calculated the spectrum of scalar
metric perturbations in the longitudinal gauge, where one finds a
scale-invariant spectrum \cite{Khoury:2001zk}. We will show in the next
section that if the collapse phase is connected to the hot big bang
expansion by a non-singular bounce then we expect the comoving
curvature perturbation, and not the curvature perturbation in the
longitudinal gauge to be conserved on large scales.

As in the pre big bang model, one requires instead a spectrum of almost
scale-invariant perturbations in an isocurvature field to lead to an
almost scale-invariant spectrum of primordial density perturbations.
As in the the pre big bang model this could be a pseudo-scalar axion
non-minimally coupled to the adiabatic field which evolves during
the ekpyrotic phase \cite{DiMarco:2002eb}. However in the ekpyrotic phase
the masses of the fields are not negligible compared with the Hubble
rate and one can also consider scale invariance due to a tachyonic
mass of an isocurvature field.

A simple example is the case of two canonical scalar fields, both
with steep negative exponential potentials \cite{Finelli:2002we,Lehners:2007ac,Buchbinder:2007ad,Creminelli:2007aq}
\begin{equation}
 \label{assisted}
 V (\varphi_1,\varphi_2) = - V_1 \exp (-\lambda_1\kappa\varphi_1) - V_2 \exp (-\lambda_2\kappa\varphi_2) \,.
\end{equation}
In slow-roll inflation it is known that a potential that is a separable sum of exponentials leads to ``assisted'' inflation \cite{Liddle:1998jc} which is a power-law expansion with power $p=\sum_I 2/\lambda_I^2$ which is larger than the power $p_I=2/\lambda_I^2$ that would be obtained for any of the fields on their own, ``assisting'' slow-roll. The same happens in ekpyrotic collapse with the potential (\ref{assisted}), although the fact that $p$ is larger than $p_I$ for a single field takes it further from the ekpyrotic limit $p\to0$.

The dynamics with multiple exponential potentials is most easily understood via a fixed rotation in field space \cite{Malik:1998gy,Koyama:2007mg}
\begin{equation}
 \phi = \frac{\lambda_2\varphi_1+\lambda_1\varphi_2}{\sqrt{\lambda_1^2+\lambda_2^2}} \,,
\quad
 \chi = \frac{\lambda_1\varphi_1-\lambda_1\varphi_2}{\sqrt{\lambda_1^2+\lambda_2^2}} \,.
\end{equation}
The potential (\ref{assisted}) is then given by
\begin{equation}
 \label{Vphichi}
 V (\phi,\chi) = U(\chi) \exp (-\lambda\kappa\phi) \,,
\end{equation}
where
\begin{equation}
 U(\chi) = - U_0 \left[ 1 + \frac{\lambda^2}{2} \kappa^2 (\chi-\chi_0)^2 + \ldots \right] \,.
\end{equation}
The ``assisted'' power-law solution corresponds to a solution where
$\phi$ evolves along the extremum, $\chi=\chi_0=$constant. One can
verify that perturbations in $\phi$ describes adiabatic field
perturbations (\ref{adiabaticpert}) along this trajectory. Thus
perturbations in $\chi$ are isocurvature perturbations described by
Eq.~(\ref{simpleKG}) with a tachyonic mass
 \begin{equation}
 m_\chi^2 = \frac{2 \kappa^2 V}{p} = - \frac{9}{2} (w^2-1) H^2 \,.
\end{equation}
The time dependent effective mass term in Eq.~(\ref{KGv}) is then
\begin{equation}
 \mu^2 = m_\chi^2 a^2 - \frac{a''}{a} = \frac{(1/4)-\nu^2}{\eta^2} \,,
\end{equation}
and the general solution is of the form given in Eq.~(\ref{Hankel}) where the order
\begin{equation}
 \nu^2 = \frac94 - \frac{18(w+1)}{(3w+1)^2} \,.
\end{equation}
Thus for $w\gg1$ we have $\nu\simeq (3/2) - (2/3w)$ and the spectral index for isocurvature field perturbations on super-Hubble scales which originate from vacuum fluctuations is given by Eq.~(\ref{ns}) with $\Delta n_{\delta\chi}\to0$ as $w\to\infty$. Note that the effective mass of the isocurvature field depends only on the combined $\lambda^2$ and not on the individual $\lambda_I$ so we do not require any cancellations between different parameters to obtain a scale-invariant spectrum; this comes out automatically for a sufficiently steep potentials with combined $\lambda\gg1$. On the other hand we have obtained an isocurvature spectrum from vacuum fluctuations only at the expense of perturbing about an unstable solution with a tachyonic direction whose mass grows proportional to the Hubble rate.

Different mechanisms have been studied which could transfer these isocurvature field perturbations to produce an almost scale invariant spectrum of scalar metric perturbations. This could occur at the bounce if the non-adiabatic field perturbations give rise to a significant non-adiabatic pressure perturbation during the bounce, or it could occur during the collapse phase as the field rolls away from the extremum $\chi=\chi_0$ either due to terms in the potential\cite{Lehners:2007ac,Buchbinder:2007ad} with $\partial V/\partial\chi\neq0$, or simply due to the tachyonic instability itself\cite{Koyama:2007ag} which naturally causes any small initial deviation from $\chi=\chi_0$ to grow. The stable late time attractor for the potential (\ref{assisted}) is an ekpyrotic collapse driven by a single field, $\phi_1$ or $\phi_2$, as in the original ekpyrotic model. Note however that isocurvature fluctuations in the other field are not then close to scale-invariant.

It is worth noting that the simple model in Eq.~(\ref{assisted}) cannot produce a red tilt, $\Delta n<0$ favoured by current obervations \cite{Dunkley:2008ie}. Thus one must introduce additional time-dependent terms either in $U(\chi)$ or breaking the exact exponential potential for $\phi$ in Eq.~(\ref{Vphichi}). One might also hope to include positive mass terms for $\chi$ to stabilise the $\chi_0$ at early times \cite{Buchbinder:2007tw}, e.g., a constant mass term which becomes dominates at early times but becomes negligible relative to the growing tachyonic mass term at late times. And eventually one must consider additional effects which will produce a transition from collapse to expansion.

\section{Perturbations through a bounce}

The calculations presented thus far are based on the dynamics of scalar fields in general relativity. This is a familiar framework for theoretical cosmology and has been thoroughly explored in the context of inflationary models of the early universe and quintessence models of the late universe. Thus the results are generally uncontroversial, although differences in approach, notably choice of gauge and conformal rescalings of the metric and/or non-minimal coupling of fields to the spacetime curvature lead to differences in the presentation of results.

However any collapse model must be connected to a expanding phase if it is to provide an explanation of initial conditions for the hot big bang cosmology, and in particular the observed spectrum of almost Gaussian, almost scale-invariant, and almost adiabatic density perturbations before the last scattering of the cosmic microwave background \cite{Dunkley:2008ie}. This is not easy. In the context of spatially flat FRW models any bounce ($H=0$, $\dot{H}>0$) requires violation of the null energy condition, $\rho+P<0$. This is not possible for any number of scalar fields with a canonical kinetic field for which $\rho+P=\dot\sigma^2\geq0$ regardless of the potential energy. Instead one requires ghost fields with negative kinetic energy which generally leads to instabilities \cite{Carroll:2003st}. The effective energy-momentum tensor of non-minimally coupled fields may violate the null energy condition \cite{Carroll:2004hc} but we have calculated our metric perturbations during collapse in the Einstein frame, and wish to set initial conditions in the primordial expanding phase where general relativity is again assumed to be valid. Therefore we will require effective violation of the null energy condition in an Einstein frame. Problems controlling the instabilities - usually by requiring some UV-completion of a low energy effective field theory containing ghosts \cite{Buchbinder:2007ad,Creminelli:2007aq,Kallosh:2007ad} - leave models invoking a bounce on much less secure foundation that other realisations of scalar fields in cosmology.

One might fear that no useful predictions can be made in the absence of a detailed physical model for the bounce. However we can make some statements about the primordial perturbations inherited from a preceding collapse phase if we make apparently reasonable assumptions about the bounce. A key assumption is that causality which limits the physical scale over which a sudden bounce can alter the scale dependence of perturbations.

We will consider first the case of isocurvature field perturbations which depend solely on the background evolution before considering the behaviour of scalar metric perturbations.

\subsection{Isocurvature field perturbations}

Let us consider the simplest case of a non-interacting scalar field
perturbation obeying the Klein-Gordon equation (\ref{simpleKG}) in
an FRW cosmology.

The effective mass, $\mu^2$ in Eq.~(\ref{mu}), contains terms from
both the physical mass, $m^2$ and the expansion, $a''/a$. If the
mass is bounded from below then all Fourier modes with wavenumber
$k\ll |am|_{\rm min}$ will follow the same evolution
$\delta\varphi_I/\delta\varphi_I(t_0)=f(t)$ independent of
wavenumber $k$. In this case the scale dependence of the spectrum of
perturbations (\ref{ns}) will not be changed. But if the spatial
gradients grow larger than the physical mass, then we need to
compare the gradients with the expansion. During collapse we related
$a''/a$ to the comoving Hubble scale, $a''/a=-(3w-1)a^2H^2$. Clearly
the Hubble length, $H^{-1}$, diverges at a bounce, but $a''$ is
finite and non-zero at a simple bounce. In fact $a''/a$ will go
through zero at some point before a smooth bounce (where $a(\eta)$
is analytic) as it is negative during the collapse phase and
positive at the bounce, and then negative again during a radiation
dominated expansion after the bounce.

Nonetheless if the bounce has a finite duration then there is always
a finite scale over which the perturbations evolution is
significantly affected by the spatial gradients, and thus a
long-wavelength regime in which the scale dependence is conserved.

In the long-wavelength limit we can model the bounce by a junction
condition obtained by integrating the Klein-Gordon equation while
imposing continuity of the field and scale factor
\begin{equation}
 \left[ \frac{v_I'}{v_I} \right]_-^+ = \left[ \frac{a'}{a} \right]_-^+ - a^2 M_I^2 \,,
\end{equation}
where we can allow for a divergent mass-term through the bounce
\begin{equation}
 M_I^2 = \lim_{\epsilon\to0} \int_{-\epsilon}^{+\epsilon} m_I^2 d\eta \,.
\end{equation}
which could lead to a scale-independent change in the perturbations.

\subsection{Adiabatic perturbations}

We can derive a generalised equation for adiabatic perturbations by
requiring that there exists a long-wavelength limit in which the
evolution of the perturbed universe is the same as that of the FRW
background \cite{Cardoso:2008gz}. The notion that long-wavelength
perturbations can be modelled as piecewise homogeneous universes is
known as the ``separate universes'' picture
\cite{Salopek:1990jq,Wands:2000dp,Bertschinger:2006aw}. This is also
sometimes called the ultra-local
approximation~\cite{Erickson:2006wc}. Once the background solution
is specified, the evolution of adiabatic perturbations on large
scales is also implicitly specified since adiabatic perturbations
are simply local perturbations forwards or backwards along this
background solution. Consistency then requires that even if a bounce
solution invokes new physics, the same new physics applies locally
to long-wavelength perturbations as applies to homogeneous FRW
cosmology.


We consider a gravitational theory where homogeneous and isotropic
spacetimes obey a Friedmann-type constraint equation, determining
the expansion rate of comoving worldlines, $\theta$, and an equation
for its evolution with respect to the proper time, $\tau$, along
these worldlines,
\begin{eqnarray}\label{perteq1}
\theta^2&=&3 f \,, \\
\frac{d}{d\tau}\theta&=&-\frac{3}{2} g \,. \label{perteq2}
\end{eqnarray}
For example, in loop quantum cosmology a modified effective
Friedmann equation (\ref{perteq1}) can be derived\cite{Singh:2006im} where $f=f(\rho)$ that leads
to a cosmological bounce, and
Cardassian models~\cite{Freese:2002sq} where $f(\rho) \propto
\rho+C\rho^n$ have been investigated. In both these examples local
energy conservation along comoving worldlines then fixes the form of
$g(\rho,p)$.

In general relativity we have $f=8\pi G\rho$ and $g=8\pi
G(\rho+P)$, where $G$ is Newton's constant. More generally, one can always
define an effective energy-momentum tensor such that the Einstein
tensor $G_{\mu\nu}=8\pi GT_{\mu\nu}^{\rm{eff}}$. {}From
Eqs.~(\ref{perteq1}) and (\ref{perteq2}) we can identify an
effective density and pressure:
\begin{equation}
 \label{eff}
\rho^{\rm{eff}} \equiv \frac{f}{8\pi G} \,, \quad p^{\rm{eff}}
\equiv \frac{g-f}{8\pi G} \,.
\end{equation}
Conservation of the Einstein tensor, $\nabla^\mu G_{\mu\nu}=0$, then
requires conservation of the effective energy-momentum tensor, which
implies
\begin{equation}
\frac{d}{d\tau}\rho^{\rm{eff}} =
-\theta(\rho^{\rm{eff}}+p^{\rm{eff}}) \,,
\end{equation}
or equivalently, from Eqs.~(\ref{perteq1}) and (\ref{perteq2}),
\begin{equation}
\frac{d}{d\tau}f = -\theta g \,.
\end{equation}
However, in the following we will allow $f$ and $g$ to be arbitrary
functions of energy, pressure or other variables.

In the linearly perturbed FRW cosmology (\ref{metric}) there is a
unit time-like vector field orthogonal to constant-$\eta$ spatial
hypersurfaces~\cite{Kodama:1985bj},
\begin{eqnarray}\label{orthovector}
N^{\mu}=\frac{1}{a}(1-A,-B_,^i) \,,
\end{eqnarray}
whose expansion rate is given by
%
%
%
\begin{eqnarray}\label{thetaeq}
\theta=3\frac{a'}{a^2}(1-A)-\frac{3}{a}\psi'+\frac{1}{a}\nabla^2\sigma
\,,
\end{eqnarray}
where a prime denotes a derivative with respect to the conformal
time $\eta$, and the anisotropic shear is
\begin{eqnarray}\label{sigma}
\nabla^2\sigma = \nabla^2(E'-B) \,.
\end{eqnarray}
At zeroth-order the shear vanishes and the background expansion rate
is $\theta_0=3{\cal H}/a$, where ${\cal H}\equiv a'/a$ is the conformal Hubble
parameter.

For the zeroth-order homogeneous (FRW) background the
equations~(\ref{perteq1}) and~(\ref{perteq2}) can be written as
\begin{eqnarray}\label{backeq1}
{\cal H}^2&=&\frac{a^2}{3} f_0 \,, \\
{\cal H}^2 - {\cal H}' &=& \frac{a^2}{2} g_0 \,. \label{backeq2}
\end{eqnarray}

We then can apply Eqs.~(\ref{perteq1}) and~(\ref{perteq2}) where we
take $f=f_0(\eta)+\delta f(\eta,{\bf x})$ and $g=g_0(\eta)+\delta
g(\eta,{\bf x})$ and the local expansion rate is given, to
first-order, by Eq.~(\ref{thetaeq}).
%
%
Neglecting all spatial gradients, we can then write the first-order
equations in terms of the lapse function $A$, its derivative, the
curvature perturbation $\psi$ and its first and second derivatives,
\begin{eqnarray}\label{perteqmetric1}
-3{\cal H}(\psi'+{\cal H}A)&=&\frac{a^2}{2}\delta f \,, \\
\psi''-{\cal H}\psi'+ {\cal H}A'+2({\cal H}'-{\cal H}^2)A &=& \frac{a^2}{2}\delta g \,.
\label{perteqmetric2}
\end{eqnarray}
Note that these equations are independent of two of the scalar
metric perturbations, $B$ and $E$ in Eq.~(\ref{metric}), which
determine the anisotropic shear (\ref{sigma}), which vanishes in
this long-wavelength limit.

For adiabatic perturbations on large scales different patches of the
inhomogeneous universe follow the same trajectory in phase space,
and the adiabatic perturbations correspond to a perturbation back or
forward with respect to this background trajectory~\cite{Wands:2000dp}. In this case the hypersurfaces of
uniform-$\theta$ and uniform-$(d\theta/d\tau)$ coincide. To
first-order this requires $\delta g/g_0'=\delta f/f_0'$.

More generally, we can write any perturbation $\delta g$ as a sum of
its adiabatic and non-adiabatic parts,
\begin{eqnarray}\label{deltag}
\delta g=\frac{g_0'}{f_0'}\delta f+\delta g_{\text{nad}} \,,
\end{eqnarray}
where $\delta g_{\text{nad}}$ is automatically gauge-invariant.
Indeed, if we identify $f$ with an effective density and $g-f$ with
an effective pressure, then $\delta g_{\rm nad}=8\pi G[\delta p^{\rm
eff}-({p_0^{\rm eff}}'/{\rho_0^{\rm eff}}')\delta \rho^{\rm
eff}]=8\pi G\delta p_{\rm nad}^{\rm eff}$. If we assume $f=f(\rho)$
in Eq.~(\ref{perteq1}) and impose energy conservation, so that
$d\rho/d\tau+\theta(\rho+p)=0$ along comoving worldlines, then we
would require from Eq.~(\ref{perteq2}) that $g=(df/d\rho)(\rho+p)$
and then $\delta g_{\rm nad}=(df/d\rho)\delta p_{\rm nad}$.

Using Eqs.~(\ref{deltag}),~(\ref{backeq1}) and~(\ref{backeq2}), we
have from Eqs.~(\ref{perteqmetric1}) and~(\ref{perteqmetric2}) that
\begin{eqnarray}\label{psiphieq}
\psi''+\frac{3{\cal H}{\cal H}'-{\cal H}''-{\cal H}^3}{{\cal H}'-{\cal H}^2}\psi'
 \hspace{0.5in} &&\nonumber\\
+\frac{{\cal H}{\cal H}'-{\cal H}^3}{{\cal H}'-{\cal H}^2}A'+\frac{2{\cal H}'^2-{\cal H}{\cal H}''}{{\cal H}'-{\cal H}^2}{A}&=&\frac{a^2}{2}\delta
g_{\text{nad}} \,.
\end{eqnarray}

Equation~(\ref{psiphieq}) includes the two gauge-dependent metric
perturbations $\psi$ and $A$. If we work in the longitudinal gauge
then we have $\Psi=\psi=A$ in the absence of any effective
anisotropic pressure\cite{Mukhanov90}. (More generally one can use the gauge freedom
to work in a pseudo-longitudinal gauge \cite{Cardoso:2008gz} which is
constructed such that $\psi=A$.) We then have
\begin{eqnarray}\label{Psieq}
\Psi''+\frac{4{\cal H}{\cal H}'-{\cal H}''-2{\cal H}^3}{{\cal H}'-{\cal H}^2}\Psi'
 \hspace{0.2in} && \nonumber\\
+\frac{2{\cal H}'^2-{\cal H}{\cal H}''}{{\cal H}'-{\cal H}^2}\Psi
 &=& \frac{a^2}{2}\delta
g_{\text{nad}} \,.
\end{eqnarray}
For adiabatic perturbations the right-hand-side vanishes and we have
a homogeneous second-order evolution equation for $\Psi$.

We can solve this equation by quadratures to find the general
solution~\cite{Polarski:1992dq,Bertschinger:2006aw}
\begin{equation}
 \label{psisolution}
 \Psi = D \frac{{\cal H}}{a^2} + C \left[-1+\frac{{\cal H}}{a^2} \int
 a^2 \,
 d\eta\right] \,,
\end{equation}
where $C$ and $D$ are constants of integration.
Although the differential equation~(\ref{Psieq}) has a singular
point when ${\cal H}'-{\cal H}^2=0$, the solution~(\ref{psisolution}) is clearly
regular through a bounce.

If we use equations (\ref{eff}) to identify an effective density and
pressure on large scales, then one can show that our generalised
perturbation equation (\ref{Psieq}) can be written in a ``general
relativistic'' form
\begin{eqnarray}
\Psi''+3(1+c_{s}^{2\rm{eff}}){\cal H}\Psi'
 \hspace{0.5in} && \nonumber\\
+[2{\cal H}'+(1+3c_s^{2\rm{eff}}){\cal H}^2]\Psi
 & =& 4\pi Ga^2\delta p_{\text{nad}}^{\rm{eff}} \,,
\end{eqnarray}
where the effective adiabatic sound speed is
\begin{equation}
c_s^{2\rm{eff}} =
\frac{p_0^{\rm{eff}\prime}}{\rho_0^{\rm{eff}\prime}} =
\frac{{\cal H}{\cal H}'+{\cal H}^3-{\cal H}''}{3{\cal H}({\cal H}'-{\cal H}^2)} \,.
\end{equation}

Our results are consistent with previous
work~\cite{Wands:2000dp,Starobinsky:2001xq} which pointed out that
the curvature perturbation on uniform-density
hypersurfaces~\cite{Bardeen:1983qw,Wands:2000dp},
\begin{equation}
\zeta \equiv - \psi - \frac{{\cal H}}{\rho'}\delta\rho \,,
\end{equation}
is conserved for adiabatic perturbations on large scales assuming
only local conservation of energy (see
also~\cite{Lyth:2003im,Lyth:2004gb,Creminelli:2004jg}).
We can define a generalization
\begin{eqnarray}\label{zeta}
\zeta_f\equiv-\psi-{\cal H}\frac{\delta f}{f_0'} \,
\end{eqnarray}
which is the gauge-invariant definition of the curvature
perturbation, $-\psi$, on uniform-expansion hypersurfaces, where
$\delta f=0$.
Using Eqs.~(\ref{backeq1}) and~(\ref{perteqmetric1}) we can write
\begin{equation}\label{zetapsi}
\zeta_f =
\frac{1}{{\cal H}'-{\cal H}^2}\left[{\cal H}\psi'+({\cal H}^2-{\cal H}')\psi+{\cal H}^2{A}\right] \,.
\end{equation}
%
%
%
In general relativity the uniform-density, uniform-expansion and comoving orthogonal
hypersurfaces coincide in the long-wavelength limit and hence
$\zeta_f=\zeta$.
Using Eqs.~(\ref{perteqmetric1}) and~(\ref{perteqmetric2}) for the
evolution of perturbations on large scales we obtain
\begin{equation}\label{zetaprime}
\zeta_f' = \frac{{\cal H}}{{\cal H}'-{\cal H}^2}\frac{a^2}{2}\delta g_{\text{nad}} \,.
\end{equation}
We see that $\zeta_f$ is constant in the large-scale limit for
modified gravitational field equations, even allowing for
non-conservation of energy, if the perturbations are adiabatic,
i.e., $\delta g_{\text{nad}}=0$ in Eq.~(\ref{deltag}).

The growing mode solution (in an expanding universe) for the
longitudinal gauge perturbation, $\Psi_+\propto C$ in
Eq.~(\ref{psisolution}), corresponds to $\zeta_f=C$, where $C$ is a
constant of integration. The decaying mode $\Psi_-=D{\cal H}/a^2$,
which dominates during ekpyrotic collapse, does not contribute to
the curvature perturbation $\zeta_f$. In a simple cosmological
bounce model, assuming a specific \textit{ansatz} for the background
evolution, one can show \cite{Cardoso:2008gz} that the growing mode
of the curvature perturbation after the bounce does not receive a
contribution from the growing mode in the collapse phase.

\section{Conclusions}

It is an intriguing possibility that the large-scale structure of
our Universe today could originate from vacuum fluctuations in a
preceding collapse phase. Cosmological models including a bounce
from collapse to expansion are certainly speculative as the
end-point of gravitational collapse remains one of the outstanding
challenges for quantum theories of gravity. We might hope that
gravitational collapse should be non-singular, but there is no
guarantee that our notions of a semi-classical spacetime will be
preserved. Loop quantum cosmology offers one framework in which
using the symmetries of FRW cosmology provides non-singular
solutions for the homogeneous background \cite{Bojowald:2001xe}, but
the dynamical evolution of an inhomogeneous universe is not known.

While the process of the bounce remains uncertain, I have argued
that assuming our semi-classical framework still holds through the
bounce, then the dynamics of a sudden bounce should affect field
perturbations above some critical scale in a scale-independent way.
And though the gravitational field equations controlling the
evolution of metric perturbations through the bounce may be unknown,
we can infer the general form of equations governing the behaviour
of metric perturbations in a long-wavelength limit in which the
inhomogeneous universe can be described locally in terms of
homogeneous patches. This shows that the comoving curvature
perturbation is constant on large scales for adiabatic
perturbations. Non-adiabatic pressure perturbations can change the
comoving curvature on large scales by changing the local equation of
state \cite{Wands:2000dp} and this could imprint the
scale-dependence of isocurvature field perturbations generated
during collapse onto the comoving curvature in the hot big bang
phase.

The decrease of the comoving Hubble length $|H|^{-1}/a$ during
accelerated collapse, like inflation in an expanding universe, leads
to sub-Hubble vacuum fluctuations producing a spectrum on
perturbations on super-Hubble scales. However there is no unique
limit in which one can obtain a scale invariant power spectrum. I
have highlighted three possibilities: (1) comoving curvature
perturbations acquire a scale-invariant spectrum in a pressureless
collapse, (2) isocurvature perturbations in axion fields in the pre
big bang scenario, or (3) isocurvature perturbations in a two-field
model of ekpyrotic collapse. In all cases we require a significant
non-adiabatic pressure perturbation to lead to a change in the
comoving curvature perturbations on super-Hubble scales either
during the collapse phase or subsequently.

In (1) and (3) we require an instability of the background solution
so that field perturbations can grow rapidly on super-Hubble scales
to keep pace with the growing Hubble rate during collapse. However
in (2) this may be avoided because the axion is non-minimally
coupled to the dilaton and so the amplitude of its vacuum
fluctuations is controlled by the Hubble rate in a conformally
related axion frame \cite{Copeland:1997ug}. The axion can acquire a
scale-invariant spectrum if the axion frame is undergoing inflation,
while in the Einstein frame the universe is collapsing.

The rapid growth of the field perturbations is likely to also lead
to the growth of second- and higher-order perturbations which could
lead to a non-Gaussian distribution of primordial density
perturbations. This has recently been calculated in multi-field
ekpyrotic
scenarios\cite{Buchbinder:2007at,Koyama:2007if,Lehners:2007wc},
case(3), and the current observational limits provide significant
constraints on the allowed parameter values. Future observational
limits should be able to effectively rule out such models as small
non-Gaussianity seems to be incompatible with these fast-roll
potentials.

In the pre big bang or ekpyrotic collapse the gravitational waves,
like the comoving curvature perturbation, acquire a steep blue
spectrum during collapse. Thus there are essentially no
gravitational waves on large scales. A detection of gravitational
waves on the Hubble scale at last scattering of the CMB would rule
out these models. On the other hand during a pressureless collapse
phase the gravitational waves acquire a scale-invariant power
spectrum, like the comoving curvature perturbation, and unlike
during slow-roll inflation, their relative amplitude is not
suppressed by slow-roll parameters. The large relative amplitude of
gravitational waves rules out this model unless the bounce phase
strongly boosts the relative scalar perturbation, which may be
possible in some cases \cite{Allen:2004vz}.


Ultimately we should should be able to use observational evidence to
confirm or rule out these very different models for the origin of
large scale structure from before the Big Bang.

\section*{Acknowledgements}

I am grateful to numerous collaborators for many discussions and collaborations upon which much of the work presented here is based, including Laura Allen, Bruce Bassett, Antonio Cardoso, Ed Copeland, Richard Easther, Kazuya Koyama, Andrew Liddle, Jim Lidsey, David Lyth, Karim Malik, Shuntaro Mizuno, Shinji Tsujikawa, Carlo Ungarelli and Filippo Vernizzi.
DW is supported by STFC.


\end{document}